\documentclass[8.5pt,twoside]{article}

\usepackage[super,sort&compress,comma]{natbib} 
\usepackage{times,mathptm}
\usepackage{sectsty}
\usepackage{balance} 
\usepackage[text={11.3cm,20.4cm},centering]{geometry} 

\usepackage{graphicx}
\usepackage{lastpage}
\usepackage[format=plain,justification=raggedright,singlelinecheck=false,font=small,labelfont=bf,labelsep=space]{caption} 
\usepackage{fancyhdr}
\begin{document}

\thispagestyle{plain}
\fancypagestyle{plain}{
\fancyhead[L]{\includegraphics[height=8pt]{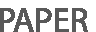}}
\fancyhead[C]{\hspace{-1cm}\includegraphics[height=15pt]{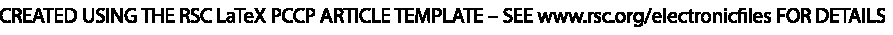}}
\fancyhead[R]{\includegraphics[height=10pt]{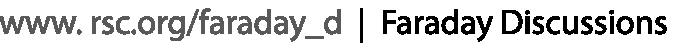}\vspace{-0.2cm}}
\renewcommand{\headrulewidth}{1pt}}
\renewcommand{\thefootnote}{\fnsymbol{footnote}}
\renewcommand\footnoterule{\vspace*{1pt}%
\hrule width 11.3cm height 0.4pt \vspace*{5pt}} 
\setcounter{secnumdepth}{5}

\makeatletter 
\renewcommand{\fnum@figure}{\textbf{Fig.~\thefigure~~}}
\def\subsubsection{\@startsection{subsubsection}{3}{10pt}{-1.25ex plus -1ex minus -.1ex}{0ex plus 0ex}{\normalsize\bf}} 
\def\paragraph{\@startsection{paragraph}{4}{10pt}{-1.25ex plus -1ex minus -.1ex}{0ex plus 0ex}{\normalsize\textit}} 
\renewcommand\@biblabel[1]{#1}            
\renewcommand\@makefntext[1]%
{\noindent\makebox[0pt][r]{\@thefnmark\,}#1}
\makeatother 
\sectionfont{\large}
\subsectionfont{\normalsize} 

\fancyfoot{}
\fancyfoot[LO,RE]{\vspace{-7pt}\includegraphics[height=8pt]{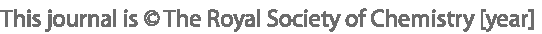}}
\fancyfoot[CO]{\vspace{-7pt}\hspace{5.9cm}\includegraphics[height=7pt]{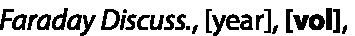}}
\fancyfoot[CE]{\vspace{-6.6pt}\hspace{-7.2cm}\includegraphics[height=7pt]{RF}}
\fancyfoot[RO]{\scriptsize{\sffamily{1--\pageref{LastPage} ~\textbar  \hspace{2pt}\thepage}}}
\fancyfoot[LE]{\scriptsize{\sffamily{\thepage~\textbar\hspace{3.3cm} 1--\pageref{LastPage}}}}
\fancyhead{}
\renewcommand{\headrulewidth}{1pt} 
\renewcommand{\footrulewidth}{1pt}
\setlength{\arrayrulewidth}{1pt}
\setlength{\columnsep}{6.5mm}
\setlength\bibsep{1pt}

\noindent\LARGE{\textbf{Cold condensation of dust in the ISM}}
\vspace{0.6cm}

\noindent\large{\textbf{Ga\"el Rouill\'e,$^{\ast}$\textit{$^{a}$} Cornelia J\"ager,\textit{$^{a}$} Serge A. Krasnokutski,\textit{$^{a}$} Melinda Krebsz,\textit{$^{b}$} and Thomas Henning\textit{$^{c}$}}}
\vspace{0.5cm}

\noindent\textit{\small{\textbf{Received Xth XXXXXXXXXX 201X, Accepted Xth XXXXXXXXX 201X\newline
First published on the web Xth XXXXXXXXXX 201X}}}

\noindent \textbf{\small{DOI: 10.1039/c000000x}}
\vspace{0.6cm}

\noindent \normalsize{The condensation of complex silicates with pyroxene and olivine composition at conditions prevailing in molecular clouds has been experimentally studied. For this purpose, molecular species comprising refractory elements were forced to accrete on cold substrates representing the cold surfaces of surviving dust grains in the interstellar medium.  The efficient formation of amorphous and homogeneous magnesium iron silicates at temperatures of about 12~K has been monitored by IR spectroscopy. The gaseous precursors of such condensation processes in the interstellar medium are formed by erosion of dust grains in supernova shock waves. In the laboratory, we have evaporated glassy silicate dust analogs and embedded the released species in neon ice matrices that have been studied spectroscopically to identify the molecular precursors of the condensing solid silicates. A sound coincidence between the 10~$\mu$m band of the interstellar silicates and the 10~$\mu$m band of the low-temperature siliceous condensates can be noted.}
\vspace{0.5cm}

\footnotetext{\textit{$^{a}$~Laboratory Astrophysics Group of the Max Planck Institute for Astronomy at the Friedrich Schiller University Jena, Institute of Solid State Physics, Helmholtzweg 3, 07743 Jena, Germany. Fax: +49-3641-9-47308; Tel: +49-3641-9-47354; E-mail: cornelia.jaeger@uni-jena.de}}
\footnotetext{\textit{$^{b}$~Institute for Geological and Geochemical Research, Research Centre for Astronomy and Earth Sciences, Hungarian Academy of Sciences, 45 Buda\"orsi street, 1112 Budapest, Hungary.}}
\footnotetext{\textit{$^{c}$~Max Planck Institute for Astronomy, K\"onigstuhl 17, 69117 Heidelberg, Germany.}}

\section{Introduction}

AGB stars and supernovae (SNs) are main cosmic dust factories. The formed stardust is finally distributed into the interstellar medium (ISM), and eventually becomes a part of cold and dense molecular clouds. Interstellar dust is exposed to destructive processes caused by supernova-induced shock waves. It was estimated that only a few percent of the total mass of stardust survive these destructive processes in the ISM.\cite{Zhukovska08,Draine09} However, observations of refractory elements in the ISM clearly show a depletion of these elements from the gas phase. Consequently, an efficient condensation process of dust grains in the ISM is required to balance the discrepancy between the stellar formation of dust grains and their interstellar destruction.\cite{Draine09} Recently,  Jones \& Nuth\cite{Jones10} discussed a compatible injection and destruction time scale for silicate dust particles, an assumption based on inherent uncertainties in the determination of destruction efficiencies.

In the ISM, refractory and other atoms and molecules generated by the erosion of grains in supernova shocks slowly accrete onto surfaces of surviving grains at low temperatures. This process is discussed to occur in the very dense cores of molecular clouds (MCs) where the temperature of dust grains is between 10 and~20 K. However, such processes may already occur in the outer, less dense regions of molecular clouds. The accreted species can finally react among each other to form solid layers on the  grains. In the very dense cores of MCs, other abundant gaseous species such as CO, 
H$_2$O, and  small carbon-rich molecules simultaneously accrete with refractory species and may prevent the formation of refractory silicate material. However, also fast desorption processes of carbon-based condensates may lead to a preferred formation of silicate condensates and to a separation of siliceous and carbonaceous dust in the ISM.  

Very recently, the condensation of SiO molecules at low temperature using neon matrix and helium droplet isolation techniques has been studied.\cite{Krasnokutski14} SiO represents the major component of the interstellar silicates. Reactions between SiO molecules were found to be barrierless. The energy of SiO polymerization reactions has been determined experimentally using a calorimetric method and theoretically with calculations based on the coupled-cluster and density functional theories. The experiments have clearly revealed the efficient formation of SiO$_x$ condensates at temperatures of about 10~K.\cite{Krasnokutski14}

In the present work, we extend our recently performed experimental studies on the cold condensation process of SiO to more complex systems containing magnesium and iron. This step is necessary to obtain more insight into the low-temperature condensation process of realistic interstellar silicates. We apply laser vaporization to silicate samples in an attempt to produce the precursor species involved in the condensation of interstellar silicates. In a first stage, the vaporized species are deposited and isolated in a Ne matrix where they are cooled down to temperatures relevant to the ISM. This stage gives us the opportunity to identify the laser-vaporized species by absorption spectroscopy in the UV and mid-IR wavelength domains. The second stage consists in annealing and warming the matrix up to 13~K until the complete evaporation of the Ne atoms in order to cause the accretion of the laser-vaporized species at low temperature.

\section{Experimental}

Chunks of two amorphous silicates, synthesized in-house by melting and quenching,\cite{Dorschner95} were used as targets. The formula of one silicate was Mg$_2$SiO$_4$, corresponding to the stoichiometry of forsterite, the Mg endmember of the olivine group. The formula of the other silicate was Mg$_{0.4}$Fe$_{0.6}$SiO$_3$, i.e., the composition of a pyroxene.

Laser vaporization was carried out using a pulsed laser source (Continuum Minilite II) emitting photons with a wavelength of 532~nm. The laser was operated with a repetition rate of 10~pulses per second. Each pulse lasted 5~ns and carried an energy of 20 to 25~mJ. The laser beam was focused at the surface of the target and, during the experiments, it was shifted every minute to vaporize a fresh part of the target and also to avoid drilling through it. Holes 0.4 to 0.5~mm in diameter were created during the experiments. Assuming this diameter coincides with the diameter of the laser beam at the target position, a fluence of 2 to 4~GW cm$^{-2}$ is inferred.

Matrix isolation spectroscopy was performed with Ne (Linde, purity 99.995\%) as the matrix material. Each matrix was grown on a KBr substrate (Korth Kristalle GmbH). This material is transmitting photons in the mid-IR wavelength domain and also at useful UV wavelengths, with a lower limit of 205~nm. A compressed-He closed-cycle cryocooler (Advanced Research Systems Inc. DE-204SL) was employed to cool down the substrate and to maintain it at low temperature. In order to form a Ne matrix, the substrate was kept at $\sim$6~K and the Ne mass flow rate was set to 5~standard cubic centimeters per minute. The target for laser vaporization and the substrate were separated by a distance of $\sim$55~mm.

The cryocooler could be rotated while being operated, allowing the substrate to face three directions and the corresponding pairs of opposite ports that equipped the vacuum chamber. These ports allowed us, in turn, to deposit the Ne atoms and laser-vaporized species on the substrate, to measure IR spectra and UV spectra in transmission. A Fourier transform IR spectrometer (Bruker VERTEX 80v) and a UV spectrometer (JASCO V-670 EX) were optically coupled to the vacuum chamber. The IR beam of the FTIR spectrometer, guided along an evacuated optical path by means of gold-coated mirrors, passed through the vacuum chamber to reach an external detector. Optical fibers fitted with collimating optics were used to carry the photons from the UV spectrometer to the vacuum chamber and back.

The IR spectra were measured by averaging 64 scans carried out at a speed of 10~kHz with a resolution of 1~cm$^{-1}$. The UV spectra were measured with equal step and resolution of 0.2~nm at a rate of $\sim$11~nm min$^{-1}$.

\section{Results}

\subsection{Analysis of the evaporated species}
The isolation of the vaporized species in the Ne matrices gives us the opportunity to identify them using absorption spectroscopy, and also to verify the presence of contaminants. For instance, our spectra show bands due to the Ne matrix-isolated contaminants CO (2141~cm$^{-1}$),\cite{Dubost76} CO$_2$ (2348~cm$^{-1}$),\cite{Wan09} H$_2$O (line systems at 1631 and 3783~cm$^{-1}$),\cite{Forney93} and (H$_2$O)$_2$ (lines at 3590 and 3734~cm$^{-1}$).\cite{Forney93}

\subsubsection{Mg$_2$SiO$_4$ target.~~} 
\begin{figure}[h]
\centering
  \includegraphics[height=6cm]{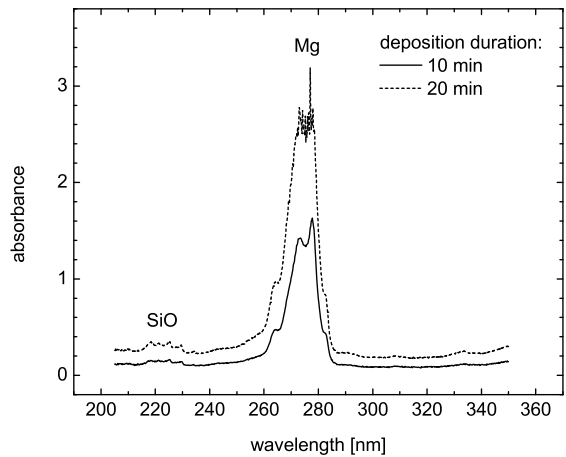}
  \caption{Ultraviolet absorption features of species isolated in a Ne matrix at 6.3~K following laser vaporization of amorphous Mg$_2$SiO$_4$. The features are attributed to SiO and Mg. The absorption line of Mg is saturated after accumulating material for 20 minutes.}
  \label{fig1}
\end{figure}

Magnesium atoms are easily detected in the UV region. In Fig.~\ref{fig1}, the line of atomic Mg at 275~nm is very strong, indicating the efficient vaporization of this element. Allowing for a matrix-induced wavelength shift, this line corresponds to that measured at 285.30~nm in vacuum.~\cite{NIST_ASD} Despite the large amount of Mg atoms, dimers are not observed. They would give rise to a band at $\sim$257~nm corresponding to the $B ^1\Pi_u \leftarrow X ^1\Sigma^+_g$ transition.\cite{McCaffrey88,Healy12} A spectrum of Mg atoms deposited in a Ne matrix to serve as a reference supports the assignment.

Still at UV wavelengths, the peaks in a very weak, structured feature originating at 234~nm correspond to bands of the $A ^1\Pi \leftarrow X ^1\Sigma^+$ transition of the SiO molecule.\cite{Hormes83,Krasnokutski14} Thus only a very little amount of this species was deposited in the Ne matrix. Accordingly, features recently attributed to SiO oligomers, which are produced by barrierless reactions between the SiO molecules, are not found.\cite{Krasnokutski14}

In the 205 to 350~nm range, Si atoms in their ground state give medium to strong absorption lines at 220.87, 243.95, and 251.51~nm in vacuum.\cite{NIST_ASD} These lines are not seen in our spectrum, even considering matrix-induced wavelength shifts. In the same region, O I atoms have only one very weak transition.

\begin{figure}[h]
\centering
\includegraphics[height=6cm]{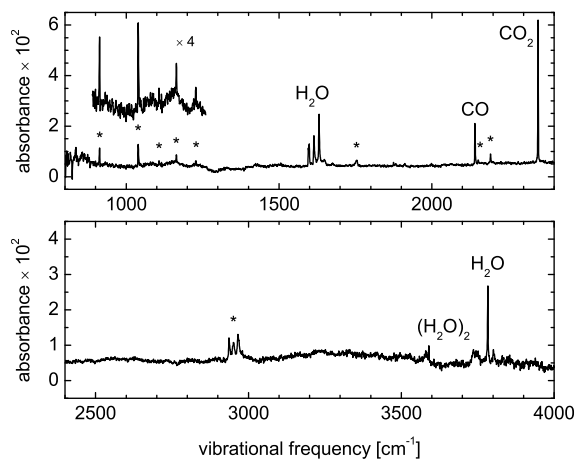}
\caption{Infrared absorption features of species isolated in a Ne matrix at 6.3~K following laser vaporization of amorphous Mg$_2$SiO$_4$. Asterisks mark the features of interest. The material had been accumulated for 20 minutes.}
\label{fig2}
\end{figure}

In the mid-IR domain, shown in Fig.~\ref{fig2}, weak absorption bands arise at 913, 1039, 1107, 1164, 1228, 1754, 2152, and 2192~cm$^{-1}$. We assign the absorption bands seen at 1039 and 1107~cm$^{-1}$ to the $\nu_3$ and $\nu_1$ mode of O$_3$, respectively.\cite{Brosset93} Another band due to O$_3$ is not seen at 2109~cm$^{-1}$ ($\nu_1$ + $\nu_3$),\cite{Brosset93} most likely because, being a combination band, it is too weak. The identification of O$_3$ could have been confirmed by the observation of the broad Hartley band in the UV spectrum, near 253~nm.\cite{Jaye98} This band, however, is not observed. The comparison between the absorption cross-sections of the UV and IR transitions would clarify our assignment. The peak at 1164~cm$^{-1}$ is attributed to O$_4^+$.\cite{Thompson89,Jacox94} Those at 913 and 2192~cm$^{-1}$ were recently reported and discussed by Jacox and Thompson.\cite{Jacox13} It was found that these absorptions may be caused by a single species, which would be a molecular complex involving H$_2$ and possibly H$_3$O$^+$ or H$_2$O$_5^+$.

The line found at 1228~cm$^{-1}$ is attributed to SiO.\cite{Khanna81,Rouille14} As expected from the analysis of the UV spectrum, the IR bands that characterize the oligomers Si$_2$O$_2$ and Si$_3$O$_3$ are not seen.\cite{Hastie69,Khanna81,Rouille14} Absorptions that would possibly reveal the presence of magnesium oxides lie at wavelengths longer than 10~$\mu$m, outside the range we have scanned.\cite{Andrews78}

Finally, a tight group of bands that arises at 2937, 2951, and 2966~cm$^{-1}$ resembles the bands of the $\nu_2$ mode of H$_2$O near 1620~cm$^{-1}$,\cite{Forney93} with a positive shift of 1337~cm$^{-1}$. The bands at 1754 and 2152~cm$^{-1}$ remain unassigned.

\subsubsection{Mg$_{0.4}$Fe$_{0.6}$SiO$_3$ target.~~} 
\begin{figure}[h]
\centering
  \includegraphics[height=6cm]{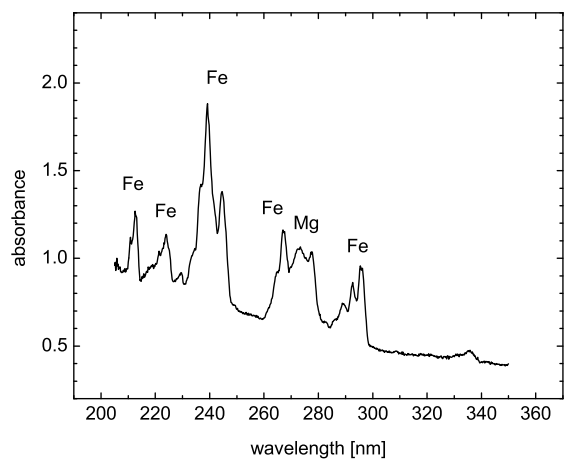}
  \caption{Ultraviolet absorption features of species isolated in a Ne matrix at 6.3~K following laser vaporization of amorphous Mg$_{0.4}$Fe$_{0.6}$SiO$_3$. The features are attributed to Fe and Mg atoms. The material had been accumulated for 20 minutes.}
  \label{fig3}
\end{figure}

In the UV region, shown in Fig.~\ref{fig3}, the lines of the Fe and Mg atoms are strong, demonstrating again the efficient vaporization of magnesium, as observed with the Mg$_2$SiO$_4$ target, and also that of iron. Only a weak contribution of SiO is possibly detected. The lines assigned to Fe I are favorably compared with the vacuum wavelengths.\cite{NIST_ASD} A spectrum of Fe atoms deposited in a Ne matrix was measured to serve as a reference. It supports the assignment. One can note that there is no evidence of the broad Hartley band of O$_3$.

\begin{figure}[h]
\centering
  \includegraphics[height=6cm]{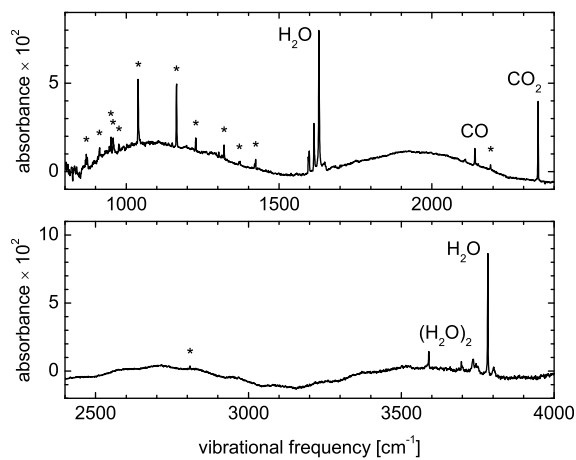}
  \caption{Infrared absorption features of species isolated in a Ne matrix at 6.3~K following laser vaporization of amorphous Mg$_{0.4}$Fe$_{0.6}$SiO$_3$. Asterisks mark the features. The material had been accumulated for 40 minutes.}
  \label{fig4}
\end{figure}

In the IR spectrum displayed in Fig.~\ref{fig4}), we find bands already observed when using the Mg$_2$SiO$_4$ target. Thus O$_3$ (1039~cm$^{-1}$), O$_4^+$ (1164~cm$^{-1}$), and SiO (1228~cm$^{-1}$) are present in the matrix. The band assigned above to the $\nu_1$ mode of O$_3$ is absent. On the other hand, new bands of O$_4^+$ isomers are detected at 1321 and 2808~cm$^{-1}$. \cite{Thompson89,Jacox94} The absorptions caused by a complex possibly involving H$_2$ and possibly H$_3$O$^+$ or H$_2$O$_5^+$ are also detected.\cite{Jacox13}

There is no evidence again of the Si$_2$O$_2$ and Si$_3$O$_3$ oligomers. The unassigned peaks found at 1754 and 2152~cm$^{-1}$ in the experiment with the Mg$_2$SiO$_4$ target are absent, like the group of bands previously observed around 2950~cm$^{-1}$.

On the other hand, there are additional bands at the positions 870, 950, 958, 976, 1369, and 1424~cm$^{-1}$. After comparison with measurements on species isolated in Ar matrices, we tentatively assign the bands at 1369 and 1424~cm$^{-1}$ to SiO$_3$ (1363.5~cm$^{-1}$ in Ar matrix) and SiO$_2$ (1416~cm$^{-1}$ in Ar matrix), respectively \cite{Tremblay96}. Other tentative assignments would be to FeO (870~cm$^{-1}$ here, 873.08~cm$^{-1}$ in Ar matrix),\cite{Green79} and OFeO (950~cm$^{-1}$ here, 945.8~cm$^{-1}$ in Ar matrix).\cite{Chertihin96}

The absorption at 976~cm$^{-1}$ has not been assigned.

\subsection{Analysis of the low-temperature condensates}
The condensates were studied by IR spectroscopy, high-resolution transmission electron microscopy (HRTEM), and energy dispersive X-ray (EDX) spectroscopy.
Generally, IR spectroscopy is the best method to follow up the silicate condensation in situ that means in dependence of the temperature. The formation of the broad silicate bands at about 10 and 20 $\mu$m corresponding to the Si-O stretching and bending modes of amorphous silicates is a clear indicator for the first appearance of a solid layer. However, due to the experimental requirements (study of the gaseous precursors and the condensate at one sample in situ), the final layer thickness of the condensed silicate  was rather small. Therefore, the typical silicate bands were weak and the measurement of the 20 $\mu$m band was not possible without having a very low signal to noise ratio.  

\begin{figure}[h]
  \centering
  \includegraphics[height=6cm]{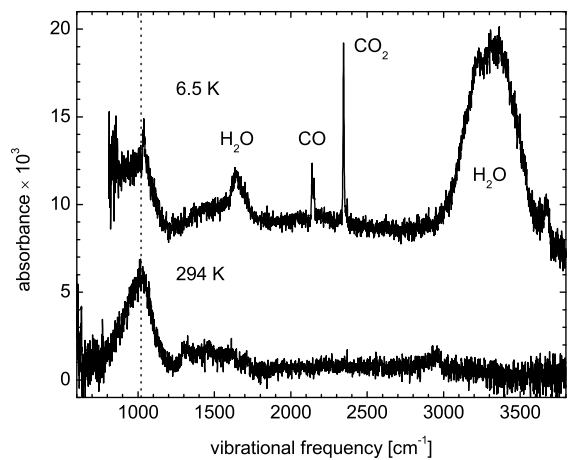}
  \caption{Infrared spectra of the condensate obtained with the Mg$_2$SiO$_4$ target. (Top) Spectrum measured at 6.5~K after warming to 13~K. (Bottom) Spectrum measured at room temperature.}
  \label{fig5}
\end{figure}

In each experiment, after annealing and complete evaporation of the Ne atoms, the species constituting the background gas have deposited on the still cold substrate giving various features. Figure~\ref{fig5} shows spectra obtained with the Mg$_2$SiO$_4$ target after evaporation of the Ne atoms and cooling to 6.5~K. Beside the absorption bands due to CO, CO$_2$, and H$_2$O, a feature is seen near 1000~cm$^{-1}$, which we attribute to a solid condensate. The profile of this band is likely affected by the presence of the broad water ice feature that peaks near 780~cm$^{-1}$.\cite{Oeberg07} After warming to room temperature, the absorption band attributed to the condensate shows an asymmetric profile, which is steeper on the higher frequency side. The maximum of the band is located at $\sim$1020~cm$^{-1}$ (9.8~$\mu$m).

Working with the Mg$_{0.4}$Fe$_{0.6}$SiO$_3$ target, we did not observe immediately the expected IR absorption near 1000~cm$^{-1}$, neither at low nor at room temperature. Nonetheless, after the experiment, a solid deposit on the substrate has been detected that has been measured using a clean KBr substrate as reference. A distinct band was observed near 1000~cm$^{-1}$ clearly pointing to the formation of a silicate material.  This finding is certainly caused by small problems with the reference measurements of pure KBr during the cooling and warming up phase. Previous temperature-dependent IR measurements of low-temperature SiO$_x$ condensates revealed the formation of the solid phase already at about 10 K.\citep{Rouille14} Similarly, the magnesium silicate condensate shown in Fig.~\ref{fig5} has either been formed at low temperature (13 K). Therefore, we act on the assumption that the solid condensate was already formed at low temperature.

\begin{figure}[h]
  \centering
  \includegraphics[height=6cm]{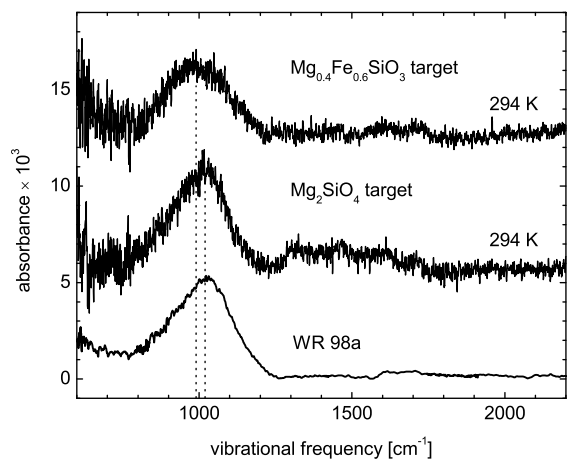}
  \caption{Infrared spectra of the final condensates obtained by laser-vaporization of amorphous Mg$_{0.4}$Fe$_{0.6}$SiO$_3$ and Mg$_2$SiO$_4$ targets measured at room temperature. The third spectrum shows the normalized silicate absorption feature toward WR 98a representing the local ISM.\cite{Chiar06} The spectra are vertically shifted for clarity.}
  \label{fig6}
\end{figure}

Figure~\ref{fig6} shows IR spectra of final condensates obtained by evaporation of a Mg$_2$SiO$_4$ and a Mg$_{0.4}$Fe$_{0.6}$SiO$_3$ target, respectively. Both spectra were taken at room temperature. A small shift of the 10 $\mu$m band is observed. The bands have their maximum at $\sim$990~cm$^{-1}$ (10.1~$\mu$m) and $\sim$1020~cm$^{-1}$ (9.8~$\mu$m), respectively. A lower polymerization degree in the condensate produced from a pyroxene-like target (Mg$_{0.4}$Fe$_{0.6}$SiO$_3$) should give a band that is shifted to smaller wavelengths compared to those produced from the olivine target (Mg$_2$SiO$_4$).  The comparison of the bands show that both condensates have to have similar stoichiometry. This has been confirmed by EDX analysis (see Table\ref{tbl:comp}).  

\begin{figure}[h]
  \centering
  \includegraphics[height=6cm]{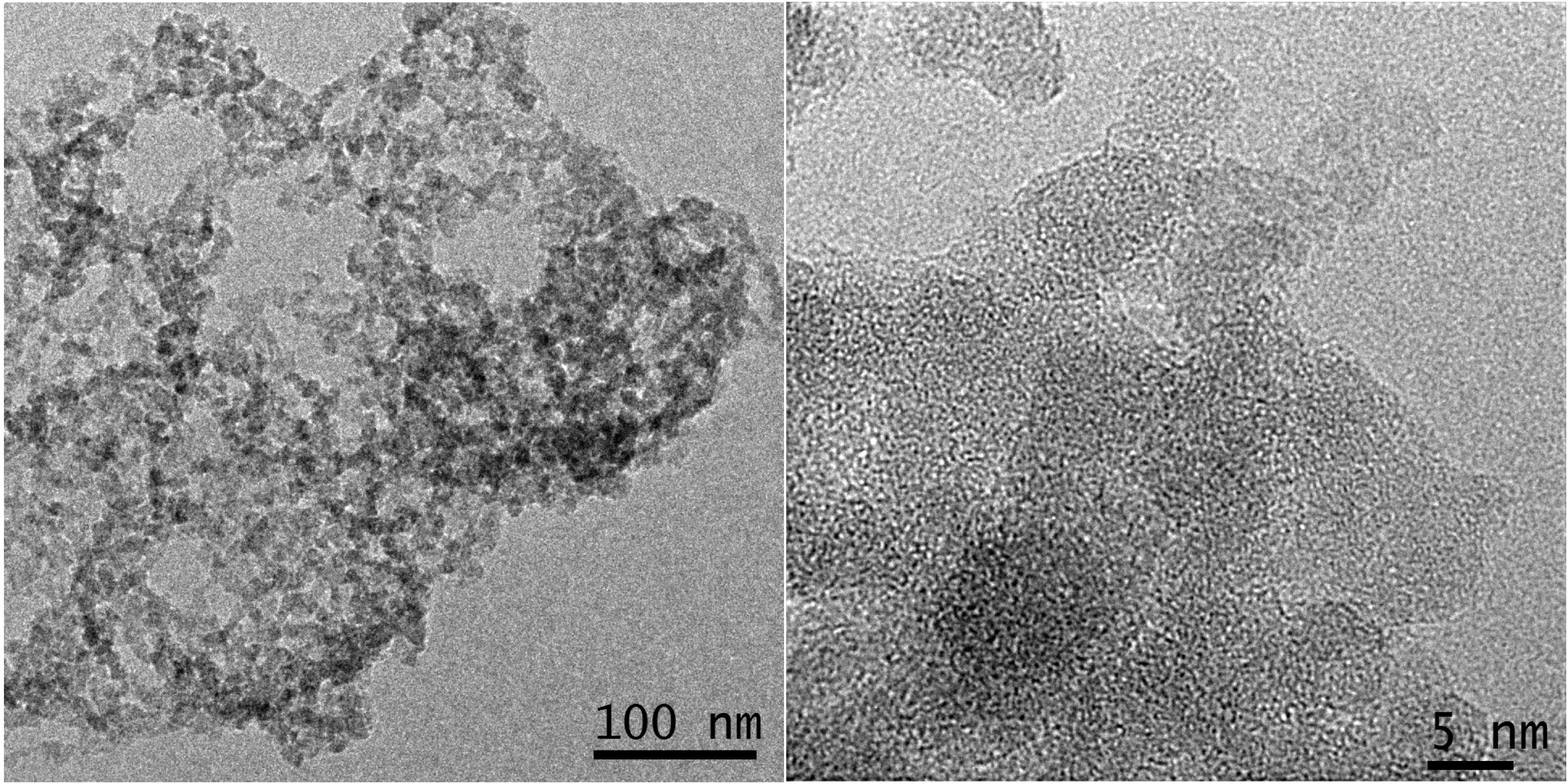}
  \caption{HRTEM image of the final condensate prepared from the evaporation and recondensation of a magnesium iron silicate at low temperatures. The left image shows an overview of a big cluster. The right image presents a direct view insight the amorphous grains. Please note the irregular structure of the interior of the grains.}
  \label{fig7}
\end{figure}

In addition, the final condensates were studied in the high-resolution transmission electron microscope (HRTEM). HRTEM micrographs of the magnesium iron silicate condensed at low temperatures show fluffy aggregates that are composed of nanometer-sized grains. The sizes of individual primary grains vary between 3 and 15~nm.  The internal grain structure is found to be completely amorphous and the final condensate shows a clear homogeneity in structure and composition (see Fig.~\ref{fig7}). No hints at phase separations of the MgFe-silicate into individual oxides such as FeO, MgO, and/or SiO$_2$ can be observed. The morphology and structure of the grains is very similar to the low-temperature condensate produced from SiO molecules.\cite{Krasnokutski14} In addition, very similar results have been obtained for the pure magnesium silicate prepared by laser ablation from a target with the composition Mg$_2$SiO$_4$. The formed fluffy aggregates are composed of small primary particles with completely amorphous structure.

The analytical characterization was complemented by the energy-dispersive X-ray spectroscopy (EDX). The method allows the determination of the final composition of the silicate material. The analysis revealed an iron-rich magnesium silicate for the condensate produced from the Mg$_{0.4}$Fe$_{0.6}$SiO$_3$ target. The composition is given in Table~\ref{tbl:comp}. Compared to the composition of the target, the material became poor in magnesium. This is probably due to a selective evaporation process. In an amorphous Mg$_{0.4}$Fe$_{0.6}$SiO$_3$ silicate, the MgO  represents the most heat-resistant component of the silicate. The composition of the final condensate produced from the  Mg$_2$SiO$_4$ target is also shown in Table~\ref{tbl:comp}. Compared to the target, the condensate is depleted in Mg and shows a pyroxene stoichiometry. 

\begin{table}[h]
\small
\centering
  \caption{~Composition of the condensed silicates by EDX analysis.}
  \label{tbl:comp}
 \begin{tabular}{l l l l l}
  \hline
   Condensate & At\% Mg & At\% Fe & At\% Si & At\% O \\
   \hline
MgFe-silicate & 4.9 & 15.8 & 17.3 & 62.0 \\
Mg-silicate & 20.8 & - & 20.9 & 58.3 \\
   \hline
\end{tabular}
\end{table}

\section{Astrophysical Discussion}

In the ISM, gaseous species may accrete on cold surfaces of pre-existing small particles. The repeated erosion of silicate stardust in SN-induced shock waves is the source of these gaseous species in the ISM. To simulate such processes in the laboratory, we have evaporated silicate dust analogs that has been produced by melting and quenching. These materials have successfully been used to model spectral energy distributions of many astrophysical sources.\cite{Gielen11,Min07,Molster02} The applied material represents a realistic source for the release of astrophysically relevant gaseous species. 

The isolation of the evaporated species in condensed rare gases cools them down before they can interact among each other. This step is necessary as laser-vaporized molecules are hot and their internal energy may affect the chemistry of accretion. The analytical characterization of the species released from silicates and finally embedded in the Ne matrix has shown that beside SiO mainly Mg, Fe, and small molecules such as O$_3$, O$_4^+$, SiO$_2$, SiO$_3$, FeO, and FeO$_2$ are precursors of the silicates. It is their accretion that resulted in the formation of silicate grains. While SiO and FeO have been discovered in the ISM,\cite{Wilson71,Walmsley02,Furuya03} the other molecules have not been reported. A model for the chemistry of silicon in dense interstellar clouds includes SiO$_2$ in its network.\cite{Herbst89} The production of SiO by interstellar shocks in molecular outflows predicts the presence of SiO$_2$.\cite{Schilke97}

The condensation of pure silicon monoxide and of more complex silicates with olivine and pyroxene composition by accretion of molecules and atoms on cold surfaces and subsequent reactions between them at temperatures between 10 and 20~K has been proven experimentally. The grains were formed by reactions of the embedded molecules during the annealing (10--12~K) and evaporation of the Ne matrix (13~K). The applied conditions are comparable to those prevailing in molecular clouds. The final condensates are fluffy aggregates consisting of small nanometer-sized primary grains. All low-temperature silicates possess amorphous structures and form fluffy aggregates.

The dust condensation at low temperature and low density in the ISM is a process that is discussed to be necessary to keep the balance between dust destruction and formation. So far, there is no exact description where the cold condensation process may take place. Turbulence in the ISM can quickly distribute the refractory elements such as Mg, Si, and other species produced in SN shocks into the surrounding medium. In SN ejecta synthesized materials are mixed by turbulence with the surrounding ISM on a time scale of 100~Myr.\cite{Oey03} One can assume a similar time scale for the mixing of elements liberated from destroyed grains. The process of cold condensation may already be active in diffuse clouds with low temperature and in low-density regions of molecular clouds. In molecular cloud environments with maximum density, where many non-refractory and refractory species and elements may accrete simultaneously on the grains, very complex ices should be formed. There are two competitive processes influencing the growth of the solid layer which is the desorption and the sticking. The desorption process is strongly determined by the bonding energy between the species and the surface. According to Draine,\cite{Draine09} for binding energies of 0.1~eV, the lifetime is about 5$\times$10$^5$~yr. Accreted species that may form strong bonds, which are typical for refractory solids, may grow very fast and remain on the surface for a long time. Diffusion and desorption processes on grain surfaces may finally trigger the formation of more stable siliceous and carbonaceous solids in addition to less stable complex ices. Furthermore, the interstellar UV field and UV photons from young stars inside the clouds, and cosmic rays can penetrate the interior of such MCs and trigger reactions between accreted molecules and clusters. 

Selection processes that may lead to the formation of spatially separated interstellar dust components have to be addressed in upcoming experimental studies.

The condensation of species produced by laser vaporization of silicates has already been studied, albeit under different conditions. The vaporized species condensed in the gas phase in a quenching gas atmosphere at high temperatures.\cite{Stephens79,Stephens95,Brucato99,Brucato02,Jaeger08,Sabri14} The particles formed in this process were deposited on various substrates before being investigated. The structure and general morphology of the high-temperature and low-temperature siliceous condensates turned out to be remarkably similar. Both are characterized by amorphous structures. In addition, both types of condensates show very similar spectral properties well comparable to the 10 $\mu$m profile of the observed interstellar silicates. So far, distinct structural and spectral differences between high- and low-temperature condensates cannot be observed.

\section{Conclusions}

The condensation of complex silicates with pyroxene composition at temperatures between 10 and 20~K by accretion of molecules and atoms on cold surfaces and subsequent reactions between them has been proven experimentally. The experiments clearly demonstrate an efficient silicate formation at low temperatures. The final condensates are fluffy aggregates consisting of small nanometer-sized primary grains. All low-temperature silicates condense in amorphous form and they were found to be homogeneous in structure and composition. To study the gaseous precursors of such condensation processes that can be formed by erosion of dust grains in SN schock waves, we have evaporated glassy silicate materials with a pulsed Nd:YAG laser. The liberated gaseous species were embedded in solid neon matrices that has been studied by spectroscopy in the UV/VIS and IR range. The IR spectral properties of low-temperature siliceous condensates do not much differ from silicates produced in high-temperature condensation processes. A sound coincidence between the 10~$\mu$m band of the interstellar silicates measured by Chiar \& Tielens\cite{Chiar06} and the 10~$\mu$m band of the low-temperature siliceous condensate can be noted.

\section*{Acknowledgements}
The authors acknowledge the support of the Deutsche Forschungsgemeinschaft through project No. He 1935/26-1 within the framework of the Priority Program 1573 "Physics of the Insterstellar Medium". M. K. is grateful for the award of an E\"otv\"os Scholarship of the Hungarian State.

\footnotesize{

\providecommand*{\mcitethebibliography}{\thebibliography}
\csname @ifundefined\endcsname{endmcitethebibliography}
{\let\endmcitethebibliography\endthebibliography}{}

}

\end{document}